\documentclass[manuscript]{aastex}

\shorttitle{Emission Height and Temperature Distribution of White-Light Flare}
\shortauthors{Watanabe et al.}

\begin{document}

\title{Emission Height and Temperature Distribution of White-Light Emission Observed by \textit{Hinode}/SOT from the 2012 January 27 X-class Solar Flare}

\author{Kyoko Watanabe\altaffilmark{1,4}, Toshifumi Shimizu\altaffilmark{1}, Satoshi Masuda\altaffilmark{2}, Kiyoshi Ichimoto\altaffilmark{3} and Masanori Ohno\altaffilmark{5}}
\email{watanabe.kyoko@isas.jaxa.jp}

\altaffiltext{1}{Institute of Space and Astronautical Science, Japan Aerospace Exploration Agency, 3-1-1 Yoshinodai, Chuo-ku, Sagamihara, Kanagawa 252-5210, Japan}
\altaffiltext{2}{Solar-Terrestrial Environment Laboratory, Nagoya University, Furo-cho, Chikusa-ku, Nagoya 464-8601, Japan}
\altaffiltext{3}{Kwasan and Hida Observatories, Kyoto University, Yamashina, Kyoto 607-8471, Japan}
\altaffiltext{4}{Research Fellow of the Japan Society for the Promotion of Science}
\altaffiltext{5}{Department of Physical Sciences, Hiroshima University, 1-3-1 Kagamiyama, Higashi-Hiroshima, Hiroshima 739-8516, Japan}

\begin{abstract}

White-light emissions were observed from an X1.7 class solar flare on 27 January 2012, using three continuum bands (red, green, and blue) of the Solar Optical Telescope (SOT) onboard the \textit{Hinode} satellite.
This event occurred near the solar limb, and so differences in locations of the various emissions are consistent with differences in heights above the photosphere of the various emission sources.
Under this interpretation, our observations are consistent with the white-light emissions occurring at the lowest levels of where the Ca {\sc ii} H emission occurs.
Moreover, the centers of the source regions of the red, green, and blue wavelengths of the white-light emissions are significantly displaced from each other, suggesting that those respective emissions are emanating from progressively lower heights in the solar atmosphere.
The temperature distribution was also calculated from the white-light data, and we found the lower-layer emission to have a higher temperature.
This indicates that high-energy particles penetrated down to near the photosphere, and deposited heat into the ambient lower layers of the atmosphere.

\end{abstract}


\section{Introduction}
\label{intro}

When strong particle acceleration occurs in solar flares, ambient coronal electrons are accelerated to very high energies of $\sim$ few tens keV to MeV.
When these accelerated electrons interact with the ambient solar atmosphere, radio and hard X-rays are produced, and apparently visible continuum (white-light) emissions are also sometimes produced.
In fact, although a small amount of white-light emission excess can be observed for many flares by high-sensitivity satellite-based and ground-based instruments, white-light emissions are mainly seen to accompany  relatively rare large flares, such as X-class flares, which generate strong particle acceleration.
Moreover, white-light emission profiles are well correlated with profiles of radio and/or hard X-rays \citep[e.g.][]{RustHegwer1975, Neidig1989, Hudson1992}.

Recent high resolution observations in white-light (by satellites and ground-based observations), hard X-rays \citep[by the \textit{Reuven Ramaty High Energy Solar Spectroscopic Imager} (\textit{RHESSI});][]{Lin2002} and radio (by ground-based radio telescopes) provide us with opportunities to study closely all of these emissions from flares.
For typical on-disk flares, the locations of these emissions are well correlated with each other in time and location \citep[e.g.][]{Metcalf2003, Xu2006, Fletcher2007}.
Since, however, on-disk flares are in essence viewed ``from above", it can be extremely difficult to determine whether the different emissions occur at the same height in the solar atmosphere, or at different heights, and there is actually an unsolved problem relating to the emission height difference between white-light and hard X-rays.

White-light emissions are generally emitted from near the photosphere.
On the other hand, non-thermal electrons, for example, of energies around $50-100\,{\rm keV}$, deposit almost all their energy when they reach the lower chromosphere.
This results in emission of hard X-rays from this location, with a total energy release that matches well with the total energy of the white-light emission \citep{Neidig1989, Ding2003, Fletcher2007, Watanabe2010}.
Observationally, the emission site of the $50-100\,{\rm keV}$ hard X-rays is estimated to be $6.5 \times 10^{3}\,{\rm km}$ above the photosphere, based on \textit{Yohkoh} observations \citep{Matsushita1992}, and $\sim 500\,{\rm km}$ from \textit{RHESSI} observations \citep[e.g.,][]{Kontar2008, BattagliaKontar2011, SaintHilaire2010}.
Even if we use \textit{RHESSI}'s results for the emission height of the hard X-ray emission, which is in the lower chromosphere, there is still a more than $500\,{\rm km}$ difference between the emission sites of those hard X-rays and the source of the white-light emissions if the white-light truly emits from the photosphere.
To reach the photosphere, electrons would have to be accelerated to greater than $900\,{\rm keV}$ \citep{Neidig1989}.
But even if such high energy electrons exist, they cannot explain the total energy of the white-light emission alone.

In contrast to the suggested photospheric height for the white-light source, \citet{BattagliaKontar2011} determined the white-light emission height to be $1,500$ to $3,000\,{\rm km}$ from \textit{Solar Dynamics Observatory} (\textit{SDO}) Helioseismic and Magnetic Imager \citep[HMI;][]{Wachter2012} continuum ($6137\,{\rm \AA}$) images.
They compared those white-light observations with hard X-ray ($25-50\,{\rm keV}$) data from \textit{RHESSI}, and found the white-light emission to be located higher than the hard X-ray emission.
From these data, they concluded that the origin of white-light emission is low energy electrons of $\le 12\,{\rm keV}$.

Using the same event as \citet{BattagliaKontar2011}, \citet{Oliveros2012} derived the emission height of hard X-rays ($30-80\,{\rm keV}$) to be about $300\,{\rm km}$ above the photosphere.
They also found the white-light ($6137\,{\rm \AA}$) emission to be located about $200\,{\rm km}$ above the photosphere, which was not significantly different from the white-light emission height.
\citet{Oliveros2012} corrected the position of the white-light continuum by fitting the limb of the full-disk HMI image via a standard inflection-point method.
They also made corrections for geocentric perspective, as \textit{SDO} was near its antisolar point at the time of the white-light observations.
That is, for this event, they found the hard X-ray source to be located very near the photosphere, and the white-light emission to be located at the same height as the hard X-ray, within the measurement uncertainties.

Following the work of \citet{Oliveros2012}, \citet{BattagliaKontar2012} corrected their alignment of the \textit{SDO} data and corrected their emission height.
Their conclusion is that the white-light emission is located $1,000\,{\rm km}$ above the photosphere, and the hard X-ray emission is located $400\,{\rm km}$ above the white-light emission.
Still there is an emission height difference.

This difference of results between \citet{BattagliaKontar2011, BattagliaKontar2012} and \citet{Oliveros2012} was due to the different methods used and assumptions made by those authors.
More precisely, the difference in the two results is due to a difference of method for aligning the different wavelength images in the two studies.
Moreover, the two studies were attempting to do critical alignment between images at different wavelengths taken by two different satellites, \textit{RHESSI} and \textit{SDO}.
If images at different wavelengths are taken by the same detector, alignment might be easier.
Similarly, alignment between images observed by different detectors can be less difficult, if both images are taken at nearly the same wavelength.
For analysis of one of the \textit{Hinode} white-light events in \citet{Watanabe2010}, we superposed a \textit{Hinode} white-light image with a \textit{RHESSI} hard X-ray image.
In doing that analysis, we first compared soft X-ray observations from the \textit{Hinode}/X-ray Telescope \citep[XRT;][]{Golub2007} with an image from a softer X-ray channel from \textit{RHESSI}.
We then coaligned an SOT/white-light image, and the XRT/soft X-ray images using the method of \citet{Shimizu2007}.
Via this process, we were able to align the \textit{Hinode} white-light and \textit{RHESSI} hard X-ray image with a high degree of confidence.

In this paper, we reconsider the relative heights of flare emissions, using observations of a white-light flare.
This flare occurred very near the west limb, and white light emission was observed in continuum bands by \textit{Hinode}/SOT.
Because it occurred near the limb, this event was ideal for estimating the source heights of white-light and hard X-ray emissions.
However, there are no \textit{RHESSI} hard X-ray data available because \textit{RHESSI} was performing annealing during the time of this flare.
Although we cannot compare the white-light emission position with hard X-rays, we will estimate the positional relationship of each multi-wavelength emission observed by \textit{Hinode}/SOT (Ca {\sc ii} H and continuum) alone.
In Section \ref{observation}, we describe the white-light data taken by \textit{Hinode}/SOT in detail.
In Sections \ref{height} and \ref{temperature}, we respectively determine the emission location of each white-light emission source and the temperature distribution of the white-light emission.
In Section \ref{discussion}, we interpret the results of the white-light emission location and temperature distribution analysis.

\section{Observations of White-Light Emission by \textit{Hinode}/SOT}
\label{observation}

The Solar Optical Telescope (SOT) of \textit{Hinode} \citep{Tsuneta2008, Suematsu2008, Shimizu2008, Ichimoto2008} has the capability to observe white-light flares.
Its broadband filter imager (BFI) can take images in red ($6684.0\,{\rm \AA}$, width $4\,{\rm \AA}$), green ($5550.5\,{\rm \AA}$, width $4\,{\rm \AA}$) and blue ($4504.5\,{\rm \AA}$, width $4\,{\rm \AA}$) continuum ranges.
SOT can also take images in the \textit{G}-band ($4305.0\,{\rm \AA}$, width $8.3\,{\rm \AA}$), formed mainly from the CH~line.
\cite{Carlsson2007} shows contribution functions for some of these filters; blue continuum is emitted from the photosphere ($\sim 0 {\rm km}$), and \textit{G}-band is also emitted from around the photosphere, but has a tail extending to about $100\,{\rm km}$ above the photosphere.

White-light flare studies with \textit{Hinode} have been performed using \textit{G}-band observations \citep{Isobe2007, Jing2008, Wang2009, Watanabe2010, Krucker2011}, mainly from the major X-class events of December 2006.
Since 2011, \textit{Hinode}/SOT has been carrying out flare observation using continuum bands during solar flares using its ``flare observation mode".
The flare observation mode is triggered by XRT, using its ability to automatically detect flares in its field of view as sudden increases in X-ray flux \citep{Kano2008}.
For this purpose, XRT periodically takes patrol images for flare detection, checking for an increase in intensity by comparing with a running-averaged patrol image.
The \textit{Hinode} satellite's Mission Data Processor (MDP), which controls XRT through sequence tables with versatile autonomous functions, monitors the level of the X-ray intensity in realtime, and raises a flare flag and determines the flaring position where the intensity exceeds a pre-determined threshold level.
This flare status and location information is sent to all three \textit{Hinode} instruments.
SOT calculates the SOT image coordinate from the location of the flare detected by XRT \citep{Tsuneta2008}.
When the flare coordinate is within SOT's FOV, SOT switches its observation mode to ``flare mode", which is a pre-determined specialized observation sequence for impulsive flares.
In flare observation mode, SOT uses a FOV of at least $108.5" \times 108.5"$ with an effective spatial resolution ($2 \times 2$ summing) of 0.108"/pixel, and takes images in at least Ca {\sc ii} H, red, green and blue filters every $20$ seconds.
Exposure times for each wavelength are $123\,{\rm ms}$, $51\,{\rm ms}$, $77\,{\rm ms}$ and $61\,{\rm ms}$, respectively.
SOT sometimes also takes \textit{G}-band, H$\alpha$, stokes V/I Na D line, and stokes IQUV Spectro-Plarimeter (SP) images.
SOT flare mode observations continue for $15$ minutes, or until the MDP flare flag drops because the X-ray intensity becomes lower than a pre-determined threshold for flare end.
By taking images in different wavelengths of continuum emission during the flare, we can observe the actual visible continuum data without any contribution from the CH~line, and we can also determine the temperature of the white-light emission in detail via the ratio of intensities from the various filters.

\begin{figure}
\epsscale{0.90}
\plotone{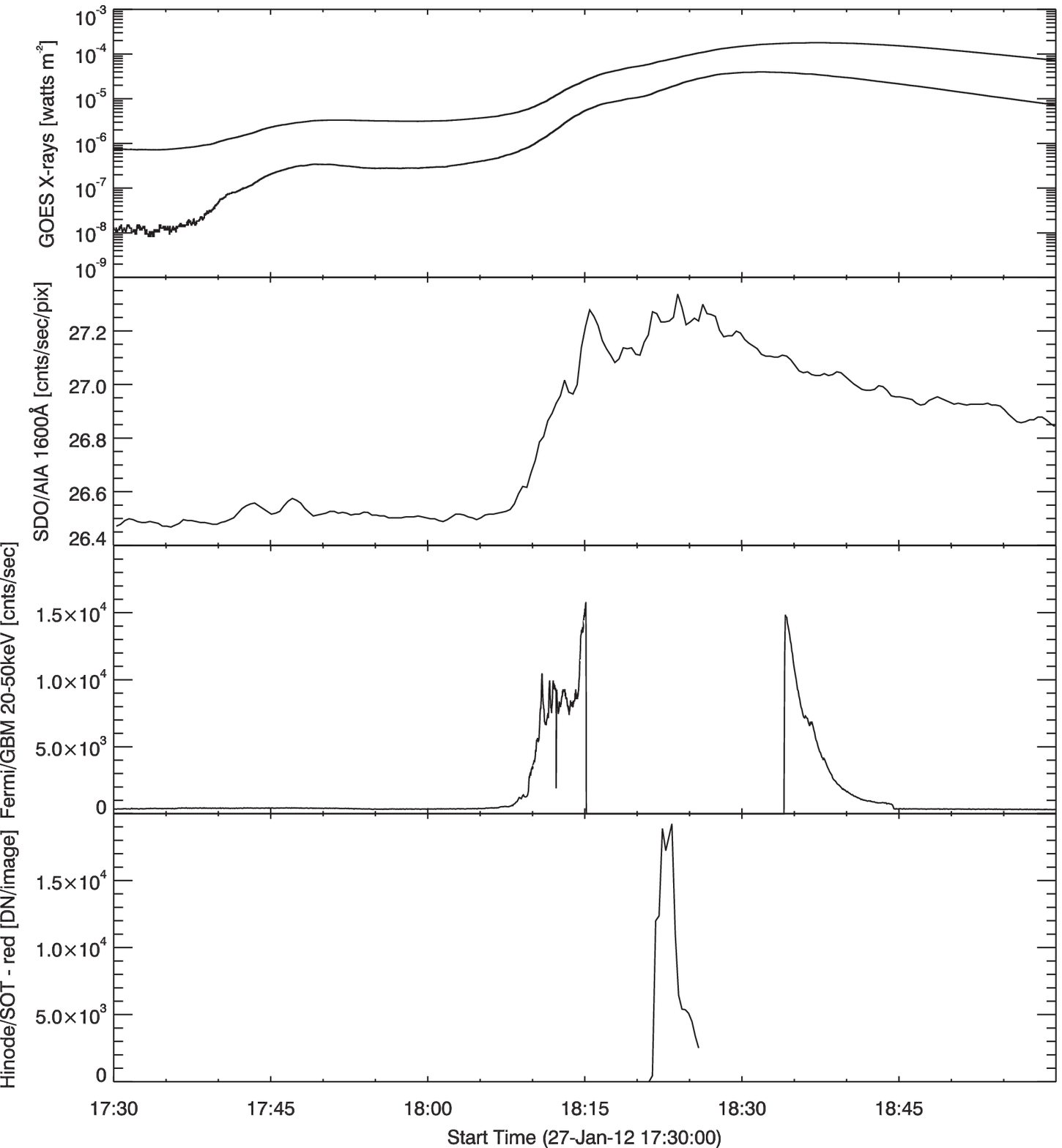}
\caption{Light curves of soft X-rays observed by \textit{GOES}, $1600\,{\rm \AA}$ UV emission by \textit{SDO}/AIA, $20-50\,\rm{keV}$ hard X-rays by \textit{Fermi}/GBM, and white-light emission by \textit{Hinode}/SOT. Over 18:15 - 18:34\,UT, there are no \textit{Fermi}/GBM data due to SAA passage. White-light (red continuum) emission is estimated  from difference images of the northeast circled region of Fig.~\ref{fig2}. The subtracted reference image for that difference was taken at 18:21:07\,UT, which was just prior to the onset of the white-light enhancements. \label{fig1}}
\end{figure}

An X1.7 class flare on 2012 January 27 was one of the events which was observed with SOT flare observation mode, and clear white-light emission was also observed.
The event occurred in active region NOAA 11402, which was located at N27W71 according to NOAA's estimation.
From the weekly report from the NOAA Space Weather Prediction Center, this flare started at 17:37\,UT, and the soft X-ray flux peaked at 18:37\,UT; this suggests that there was an 1 hour period between the start and maximum of the X-ray flux.
Upon closer inspection however, the main phase of the X1.7 flare did not start at 17:37\,UT.
The top panel of Fig.~\ref{fig1} shows the soft X-ray profile of this event.
It appears as if a C-class flare occurred at 17:37\,UT, and during the decay phase of that leading flare, the main phase of the X-class flare started around 18:10\,UT.
The same impulsive behavior at about 18:10\,UT occurs in the light curve of the UV emission of \textit{SDO}/AIA $1600\,{\rm \AA}$  as shown in the second panel in Fig.~\ref{fig1}.
During this flare, unfortunately, there were no \textit{RHESSI} hard X-ray data due to RHESSI's maintenance (annealing).
The \textit{Fermi}/GBM instrument however was operating, and detected substantial hard X-ray emissions, including at $\ge 100\,{\rm keV}$ from 18:08\,UT.
The third panel of Fig.~\ref{fig1} shows a light curve of $20-50\,{\rm keV}$ hard X-rays observed by \textit{Fermi}/GBM.
Because \textit{Fermi} entered the South Atlantic Anomaly (SAA) at 18:15\,UT, hard X-ray data are not available after this time.
However, the hard X-ray profile is closely correlated with the rising phase of the \textit{GOES} soft X-ray profile up to the M-class level, and with the impulsive increase of AIA $1600\,{\rm \AA}$ emission.
Moreover, the $20-50\,{\rm keV}$ GBM emission shows another burst just before entering the SAA.
We are confident that the rapid increase of $20-50\,{\rm keV}$ emission after 18:14\,UT is not the effect of radiation from SAA, as the emission is too strong compared to the ordinary SAA response of \textit{Fermi}/GBM.
Because this 18:14\,UT hard X-ray burst is correlated with the burst of $1600\,{\rm \AA}$ UV emission, it seems that another new burst occurred after 18:14\,UT.
However, this burst did not lead to increased X-ray emission and did not produce any white-light emission.
The soft X-ray flux surpassed the M5-level at 18:19\,UT, and also increased to X-class shortly after 18:20\,UT.
From these data, we can determine that the main impulsive phase of the X1.7 flare started just after 18:20\,UT.
Moreover, clear white-light enhancement was seen after 18:21\,UT, as shown in the bottom panel of Fig.~\ref{fig1}.

\begin{figure}
\epsscale{0.90}
\plotone{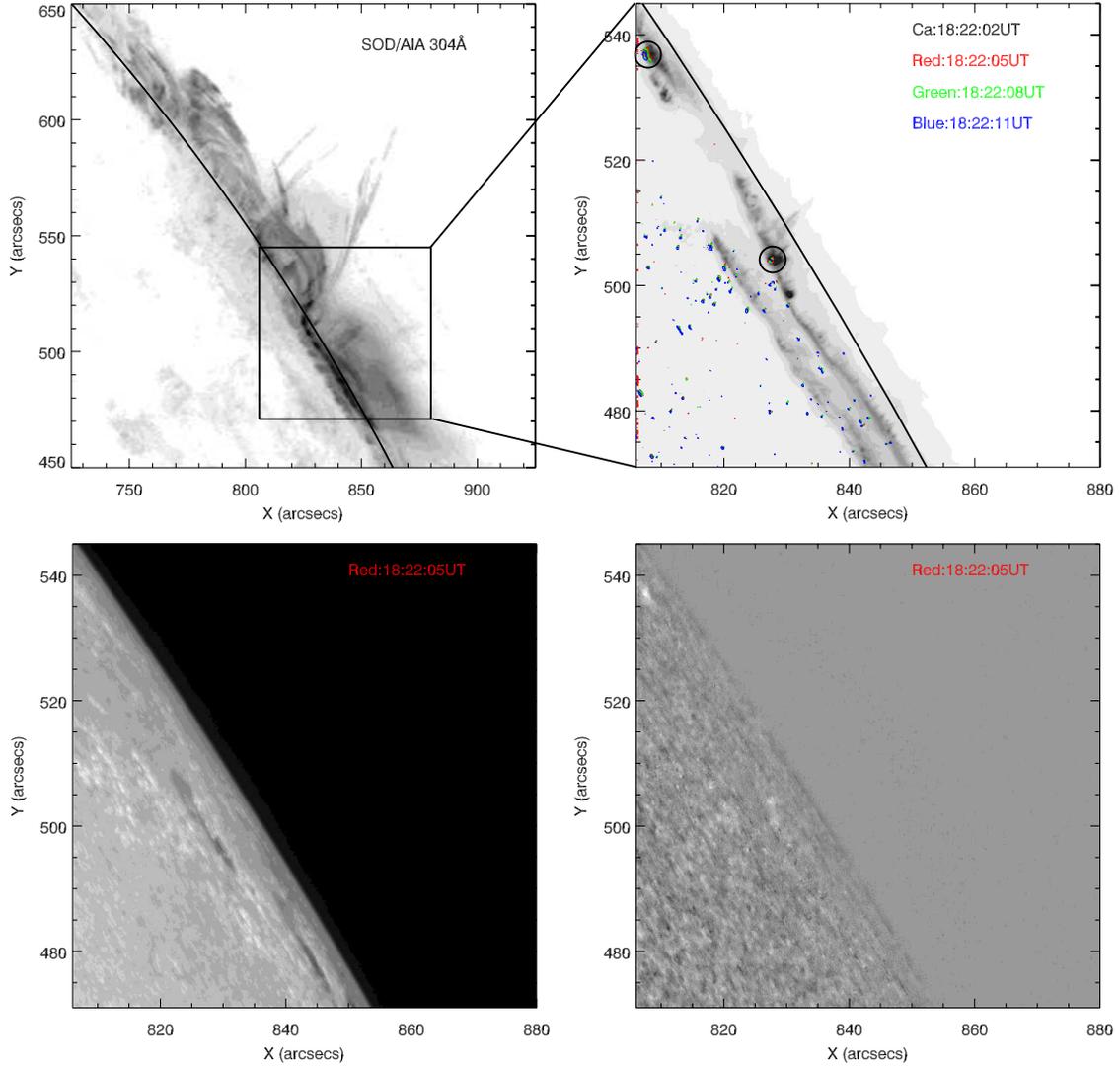}
\caption{\textit{Top-left:} \textit{SDO}/AIA $304\,{\rm \AA}$ image taken at 18:22:08\,UT on 2012 Jan 27. The box at the center of the image shows the FOV of the images in the three other panels. The black curve is the solar limb. \textit{Top-right:} Observed Ca {\sc ii} H emission with white-light emission from \textit{Hinode}/SOT. The overlaying red, green and blue contours indicate intensity deviations $\ge 3\sigma$ above background for each wavelength channel. Significant white-light emission is mainly seen in the circled positions. The Ca {\sc ii} H images was taken at 18:22:02\,UT, and the red green and blue images were taken at 18:22:05, 18:22:08, and 18:22:11\,UT, respectively. The black line is again the solar limb. \textit{Bottom-left:} Observed red continuum image at 18:22:05\,UT. \textit{Bottom-right:} A difference image formed by subtracting a reference image at 18:21:07\,UT from the image in the lower left panel. This image was used for the intensity-significance determination for the contours shown in the top-right panel. \label{fig2}}
\end{figure}

At 18:17\,UT on January 27, SOT switched to flare mode, in time for observations of the main phase of this X1.7 flare.
Although ribbons of Ca {\sc ii} H emission were already formed at the start time of flare mode observation, continuum emission was not seen at this time.
By taking the differences of continuum images, continuum enhancements could be clearly seen after 18:22\,UT.
The top-right panel in Fig.~\ref{fig2} shows an Ca {\sc ii} H image from 18:22:02\,UT; this is the Ca {\sc ii} H image nearest the time that the continuum emission was first clearly observed, and the overlaid contours show the observed continuum emissions (red, green and blue).
These color contours indicate $3\sigma$ deviation above the background of the subtracted image.
An example of a background-subtracted (i.e. difference) red image is shown in the bottom-right panel in Fig.~\ref{fig2}, and the original (non-differenced) image is shown in the bottom-left panel in Fig.~\ref{fig2}.
The reference images used for the fixed difference images were taken at 18:21:07\,UT for the red images, 18:21:11\,UT for the green images, and 18:21:14\,UT for the blue images.

For aligning the images prior to subtraction, gaps of sub-pixel size and the difference of the plate scale between the continuum and Ca {\sc ii} H images are corrected as described in \citet{Shimizu2007}.
Accuracy of the co-alignment is better than 1 pixel ($\sim70\,{\rm km}$).
The continuum emissions are seen at about the same locations where very strong, almost saturated, Ca {\sc ii} H emission was observed.
White-light emissions were mainly seen on the northeast Ca {\sc ii} H ribbons and in the middle of the west Ca {\sc ii} H ribbon, as circled in the top-right panel of Fig.~\ref{fig2}.
These white-light and Ca {\sc ii} H emissions are almost correlated with the footpoints of the flaring loop, based on comparisons with \textit{SDO}/AIA $304\,{\rm \AA}$ images such as that shown in the top-left panel in Fig.~\ref{fig2}.
The strongest white-light emission was seen around the east edge on the image that is located on the northeast Ca {\sc ii} H ribbons.
The white-light source at the middle of the west Ca {\sc ii} H ribbon is not as strong as the northeast white-light source, and its emissions always appear in patches, which makes it difficult to estimate the emission location and this heights.
Therefore, in this paper, we analyze only the northeast white-light source to determine the respective white-light emission heights above the solar limb.
The location on the Sun of this white-light emission was estimated to be N33W83 in the \textit{Hinode} image, so this event occurred very near the limb.

\section{Location of White-Light Emission}
\label{height}

\begin{figure}
\epsscale{1.00}
\plotone{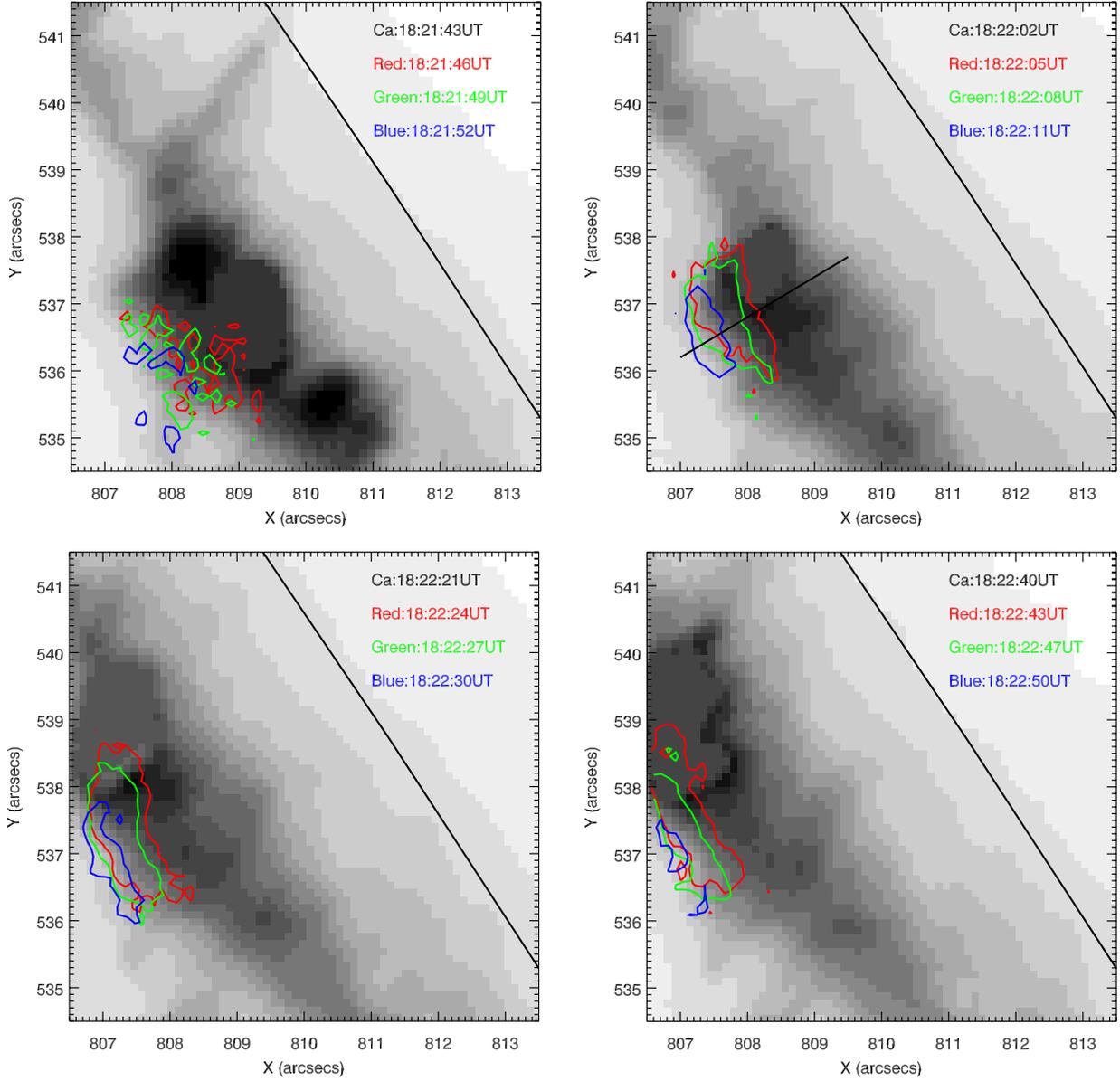}
\caption{Observed white-light flare emission in these wavelength channels (red, green, blue) overlaid onto a Ca {\sc ii} H image. The red, green and blue contours indicate intensities $\ge 3\sigma$ above the background for each channel. The times of the Ca {\sc ii} H images are shown at the top of each image, and the times of the white-light emissions are also shown on each image. The black line in each panel is the solar limb.\label{fig3}}
\end{figure}

Fig.~\ref{fig3} shows a Ca {\sc ii} H image of a closeup of the region around the northeast white-light emission location, with contours overlaid showing $3\sigma$-level deviations above the continuum emissions.
Before 18:22\,UT, white-light emission was not clearly seen, as shown in the top-left panel of Fig.~\ref{fig3}.
Only a few pixels exceeded $3 \sigma$ above the background, and moreover any excess occurred in disjoint patches.
After 18:22\,UT, continuum emission was clearly observed, as shown in the top-right panel of Fig.~\ref{fig3}, and continued to emit for 1 minute following the onset of strong Ca {\sc ii} H emission, as shown in the bottom panels of Fig.~\ref{fig3}.
These continuum emissions were always located on the disk-center side of the Ca {\sc ii} H emission, and spread over $\sim 1 \,{\rm arcsec}$ size.

When we examine the location of the continuum emission in the separate wavelength channels in detail, it is seen that each source of emission is not located at the same location; rather, the emission sources are significantly offset in the same direction.
Usually, the green emission is shifted to the southeast of the red emission location, and the blue emission is also shifted in the same direction from the green emission.
In order to estimate the location of each continuum emission in detail, we put a fiducial marker on the images, as shown by the black line in the top-right panel of Fig.~\ref{fig3}, and calculate the relative position of each emission source along that fiducial.

Fig.~\ref{fig4} shows the relative position of Ca {\sc ii} H, red, green and blue continuum emission along the fiducial line in the top-right panel in Fig.~\ref{fig3}.
Fig.~\ref{fig4} shows that the peak red emission is located more than $300\,{\rm km}$ from the peak of the Ca {\sc ii} H emission.
Also, the peak locations of the red and green emissions differ by more than $100\,{\rm km}$, and those of the red and blue emissions differ by about $400\,{\rm km}$.

The differences in these locations are significant by virtue of the $0.2\,{\rm arcsec}$ resolution of our SOT images.
Such differences in emission location have not previously been described from observations of on-disk flares.
For example, for the \textit{Hinode} white-light observations reported by \citet{Krucker2011} and \citet{Watanabe2010}, the location of the white-light emission channel and the hard X-ray emission were well correlated, and they also had the same structure.
Because this flare occurred almost at the limb, we suspect that these location differences are due to differences of emission height at each wavelength.
We will discuss this possibility further in Section~\ref{discussion}.

\section{Temperature Distribution of White-Light Emission}
\label{temperature}

From SOT continuum data, we can determine the temperature of the white-light emission in detail, because we observed the white-light emissions using three different-wavelength continuum bands.
To calculate the temperature of the white-light emission, we normalize the emissions from the three wavelength bands to a quiet sun fit to a $5780\,{\rm K}$ blackbody spectrum \citep{Cox2000}.
For this normalization, we used synoptic white-light data taken at disk center at 18:05\,UT on 2012 January 27, and normalized the observed intensity (in DN units) to fit the $5780\,{\rm K}$ blackbody spectrum, as shown in the top panel in Fig.~\ref{fig5}.
Using the normalized emission factors for the three continuum bands from synoptic data, we then fit the three normalized data points for the other data using the Planck formula ($B_{\lambda}(T) = 2hc^2/\lambda^5 (e^{hc/\lambda kT}-1)$), and estimated the temperature of the observed white-light emissions.
A fitting result for the white-light emission observed at the $\times$ mark in the top-right panel in Fig.~\ref{fig6} is shown in the bottom panel in Fig.~\ref{fig5}.
For this example, the observed three wavelength band emissions are well fitted to the Planck function, and the temperature for this pixel is estimated to be $4880\,{\rm K}$; this is the standard temperature of the quiet sun around the W83 position.

Fig.~\ref{fig6} shows the temperature distribution of the observed white-light emission for four different sets of observation times.
The temperature drop near the limb is due to limb darkening.
High temperature regions in the top-left panel in Fig.~\ref{fig6} are faculae (with no heating yet from the flare).
Heating due to the white-light emission amongst these faculae appears as shown in the top-right and bottom panels of Fig.~\ref{fig6}.
The average temperature around the locations showing continuum enhancements is about $5000\,{\rm K}$.
The temperature of white-light emission was estimated as $\sim 10,000\,{\rm K}$ in previous studies \citep[e.g.][]{Kretzschmar2011}, however, a much lower temperature was obtained in this study.
There is a possibility that the lower temperatures we find here for white-light emission are due to limb darkening, because this event occurred very near the limb.
Next, we investigate the temperature distribution around the emission region for each color in detail.

Table~\ref{tbl1} shows the averaged temperature of the white-light emission for each of the wavelength regions.
These are averaged in the sense that, for each wavelength channel, we averaged the temperature over all of the regions of $\ge 3 \sigma$ intensity for that wavelength.
The calculated temperatures over 18:21:46$-$18:21:52\,UT were less than $4800\,{\rm K}$ for the green and blue channels, and moreover there is no obvious trend of the temperatures with height.
That is, from highest to lowest height in the atmosphere the wavelength regions are ordered from red, to green, to blue, but the temperatures do not follow a monotonic increasing or decreasing trend correlated with these respective heights.
For the following time period of 18:22:05$-$18:22:11\,UT, corresponding to the upper-right panel of Fig.~\ref{fig6}, there is not yet a clear trend in the temperatures with height, although the temperature values are close together compared to the first time period, and they are substantially hotter ($\ge 100\,{\rm K}$) than in the first time period.

By the time of the third (18:22:24$-$18:22:30\,UT) and fourth (18:22:43$-$18:22:50\,UT) time periods, the temperatures have again increased over the earlier time periods, and now there is a clear trend, with the blue, green, and red regions progressively showing hotter to cooler temperatures.
That is, the temperatures for these two time periods show a trend of hotter to cooler as the height goes from lower to higher in the observed region of the photosphere.
Considering that the error in these temperatures is about $35\,{\rm K}$, the temperature trend between the red and green regions changes significantly and has a clear trend with height by the time of the second to forth time periods.
Although the temperature change between the green and blue regions is within the error range by the time of the second to forth time periods, we can say that there is a weak temperature trend between the green and blue regions by the time of the third and forth time periods.
We conclude that for the first time periods, when there was still little flux from the flare at white-light wavelengths, the obtained temperatures show a more-or-less random variation among the three wavelength channels.
After the white-light flux from the flare increased (especially during the two latest time periods of Table~\ref{tbl1}), the monotonic tread of decreasing temperature with decreasing height became apparent.

\section{Discussion}
\label{discussion}

\begin{figure}
\epsscale{1.00}
\plotone{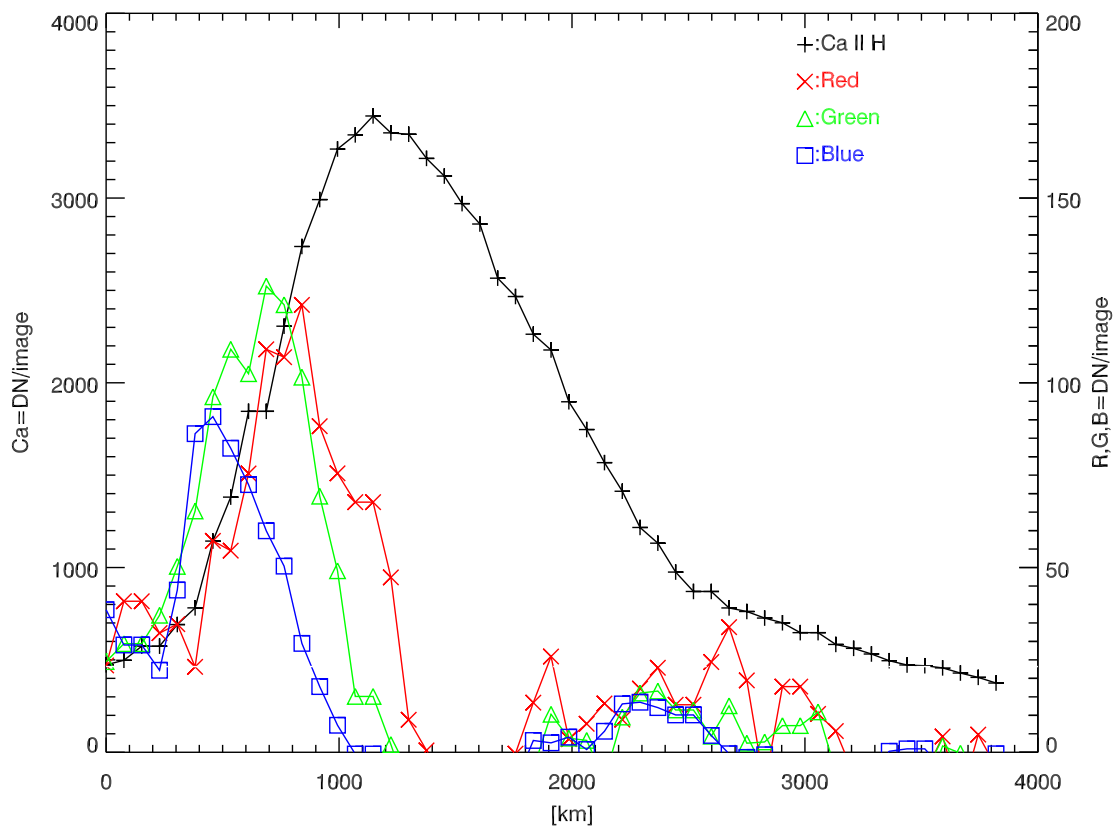}
\caption{Relative emission locations of red, green and blue-channel white-light emissions and Ca {\sc ii} H emission. The X-axis shows height from the solar surface, where we have compensated for the $7$-degree longitudinal offset from the limb of the flare. This compensation results in the photosphere (height = $0\,{\rm km}$) being located at X=$500\,{\rm km}$ in this plot. Continuum signals at $\>1500\,{\rm km}$ are less than $\sim 1 \sigma$ level of the emission almost everywhere. \label{fig4}}
\end{figure}

\begin{figure}
\epsscale{0.70}
\plotone{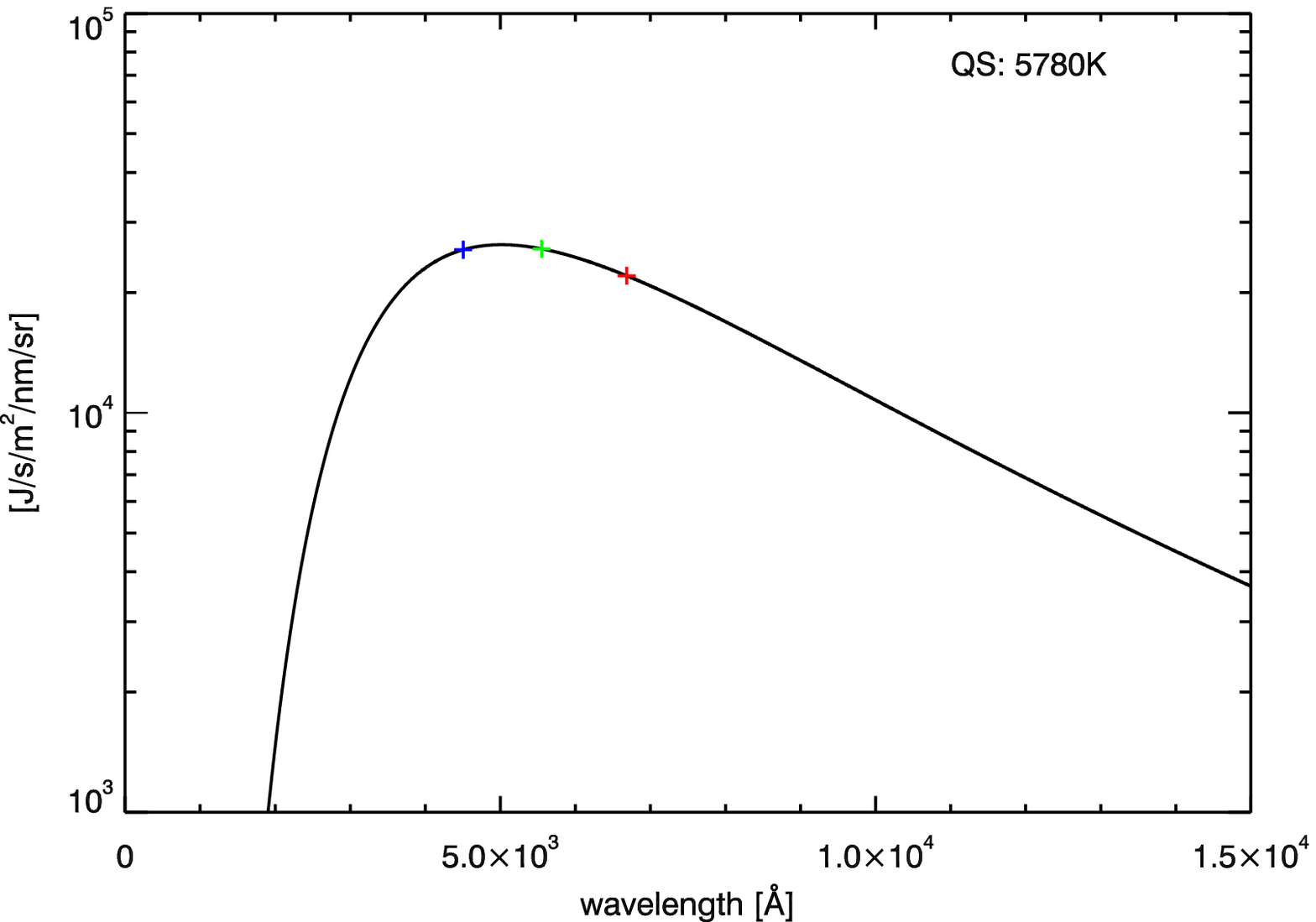}
\plotone{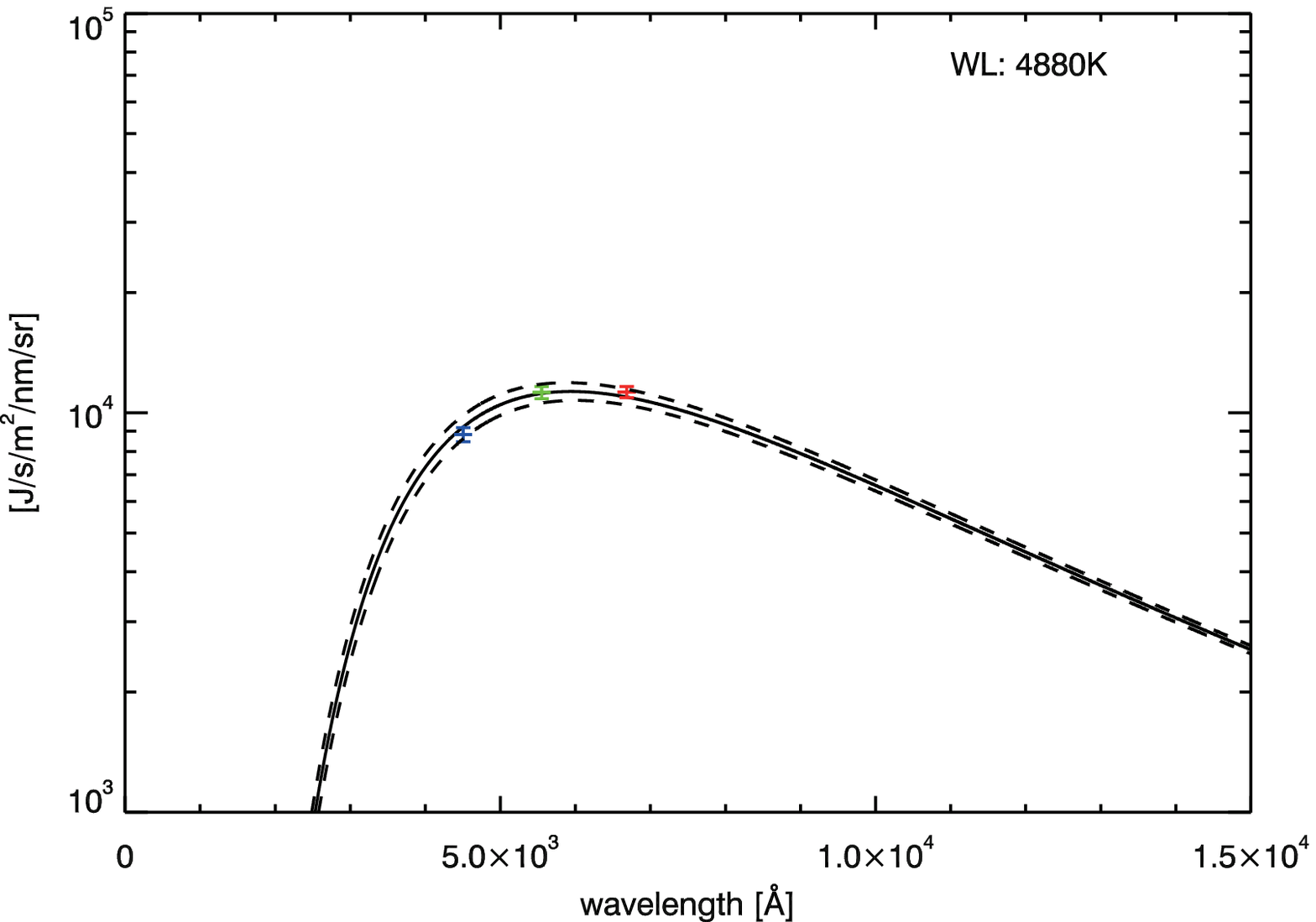}
\caption{Results of fitting observed white-light emissions to the Planck function. \textit{Top:} The three-band white-light emissions in the quiet sun normalized to fit a $5780\,{\rm K}$ blackbody spectrum. \textit{Bottom:} A temperature is estimated by fitting the the Planck function to the normalized data. For the pixel given by the $\times$ mark in the top-right panel Fig.~\ref{fig6}, the temperature is estimated as $4880\,{\rm K}$. Errors are $1 \sigma$ deviations of the observed DN for each pixel. Dashed lines indicate Planck functions at $\pm50\,{\rm K}$ from $4880\,{\rm K}$. \label{fig5}}
\end{figure}

From our analysis of \textit{Hinode}/SOT observations of the 2012 January 27 white-light flare, we found different source emission locations for each of the three channels of white-light emission that were observed.
Because the location on the Sun of this flare seen from \textit{Hinode} was N33W83, there is a possibility that this location difference was due to differences in the height.
In order to translate the observed distances into vertical heights above the photosphere, we accounted for the $7$-degree longitudinal rotation to the limb in Fig.~\ref{fig4}.
Considering that the horizontal axis in Fig.~\ref{fig4} expresses vertical height above the photosphere, the red, green and blue emission is located below the Ca {\sc ii} H emission.

From these SOT data alone, this ordering of the heights of the Ca {\sc ii} H, and the blue, green, and red continuum emission, is all we can determine about the emission heights.
We cannot, for example, determine with these data the height of the Ca {\sc ii} H emission in the solar atmosphere, say, relative to the $5000\,{\rm \AA}$ $\tau=1$  photosphere. 
Other workers however have investigated this question.  
\citet{Carlsson2007} and \citet{JudgeCarlsson2010}  for example estimated the Ca {\sc ii} H emission to be at $800\,{\rm km}$ above the $\tau=1$ photosphere, which is where it is normally defined at $5000\,{\rm \AA}$, and according to \citet{JudgeCarlsson2010}, the continuum at the limb forms about $375\,{\rm km}$ higher than at disk center \citep[e.g.,][]{Athay1976} and the Ca {\sc ii} H limb lies $450 \pm 34\,{\rm km}$ above the blue limb \citep{Bjolseth2008}.
Although we do not know the height of the lower edge of the Ca {\sc ii} H emission from this information, the Ca {\sc ii} H limb at the quiet sun forms near heights of $825\,{\rm km}$, which is already the upper photosphere.
Although we observed that the Ca {\sc ii} H ribbons during this flare were located higher in the atmosphere, as shown in Fig.~\ref{fig4}, this is due to the high density plasma of the solar flare from the corona.
The response of the Ca {\sc ii} H emission extends down to the lower photosphere, so the edge of the Ca {\sc ii} H emission might be coming from the photosphere.
From this investigation, the lower edge of the Ca {\sc ii} H emission, that is the same as the height of the peak emission of the blue band (about $500\,{\rm km}$ in Fig.~\ref{fig4}), might be coming from the photosphere.
Thus, it is confirmed that the observed continuum emission was not located in the chromosphere as in the \citet{BattagliaKontar2011}'s event, but instead the white-light emission was emitted from the photosphere.

We found that higher temperatures were observed in the blue region than the red region.
As discussed above, because this flare was located very near the limb, we can consider that the red region is higher in the atmosphere than the blue layer.
So, this temperature difference means that the lower layer has a higher temperature at the footpoint of the flare loop.
This result indicates that the source of the white-light flare emission (heat source) exists at low heights in the photosphere of the flaring loops, and the higher photospheric layers of the flare loops are heated from the lower layers of the photosphere.
There is no evidence of heating from the lower chromosphere or from thicker layers above the photosphere.
It seems that accelerated particles directly penetrate to the lower layers of the photosphere, and heat the lower atmosphere more than the higher atmosphere.
The observed flare hard X-ray source of \citet{Oliveros2012} suggests the existence of accelerated electrons in the photosphere, which is consistent with what we observe here.

\begin{figure}
\epsscale{1.00}
\plotone{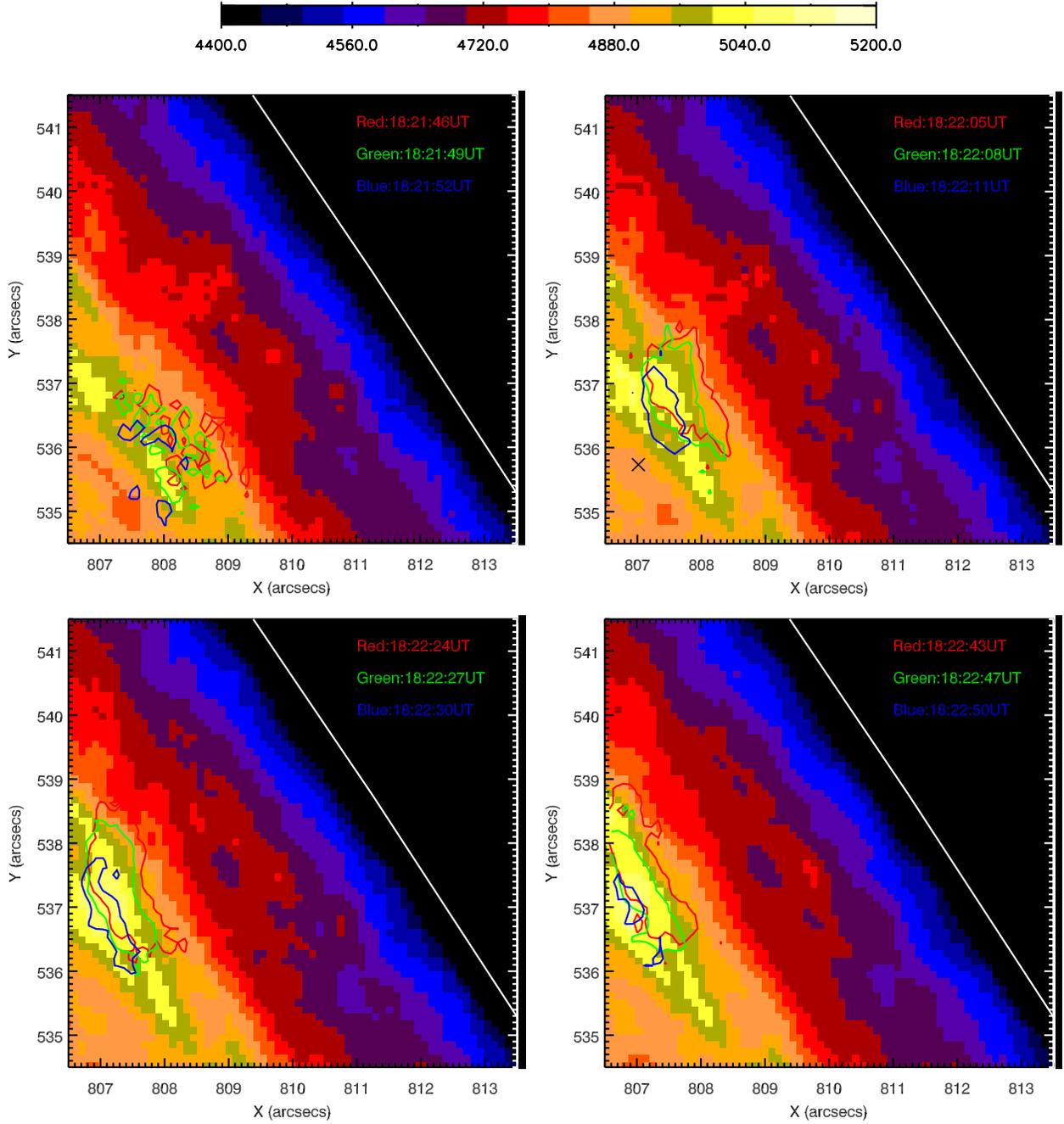}
\caption{Temperature distribution of white-light emission. The same color contours for the three continuum channels as in Fig.~\ref{fig3} are overlaid. The white line in each panel is the solar limb. The ``$\times$" mark in the top-right panel indicates the pixel for which the temperature was calculated in the bottom panel of Fig.~\ref{fig5}. \label{fig6}}
\end{figure}

The temperature of the observed white-light emissions were estimated as $\sim 5000\,{\rm K}$; this is a much lower temperature than previous studies \citep[e.g.][]{Kretzschmar2011}.
One effect that reduces the temperature is limb-darkening.
The diminishing of intensity at the limb from the disk center occurs at visible wavelengths, and this diminishing results in a temperature drop.
In fact, the estimated temperature of the quiet sun in Fig.~\ref{fig6} is about $4900\,{\rm K}$, which is a very low temperature compared with the usual temperature of the quiet Sun ($5780\,{\rm K}$).
The form of the limb darkening is given by $I(\theta) = I(0) [1-u(1-\cos \theta)]$.
Here $\theta$ is angle from the disk center, and $I(0)$ and $I(\theta)$ are the intensity at disk center and at location $\theta$, respectively.
Also, $u$ is the limb darkening coefficient; e.g. $u=0.56$ at $6000\,{\rm \AA}$, and $u=0.95$ at $3200\,{\rm \AA}$.
Using $u=0.56$, at W83, $I(\theta)$ for all the white-light emission observed by \textit{Hinode}, is about half the intensity at disk center.
Because the intensity is proportional to the fourth power of the temperature (Stefan-Boltzmann law), $5780\,{\rm K}$ (quiet sun temperature at disk center) will be $\sim 4900\,{\rm K}$ around W83.
This estimate is correlated with the estimated temperature shown in Fig,~\ref{fig6}.
Using this limb darkening effect, we can estimate the temperature of the white-light emission at the disk center, and find it to be almost $6000\,{\rm K}$, based on $\sim 5000\,{\rm K}$ being the observed temperature at W83.
This is still lower than the standard $6000\,{\rm K}$ temperature of the photosphere, and is almost the same as the temperature of the non-flaring photosphere.

In fact, the temperature increases by only $100-200\,{\rm K}$, and the required energy for this heating is about $10^{25 \sim 26}\,{\rm erg/s}$ - which is not so big.
To study whether this energy is consistent with the hard X-ray energy from accelerated particles or not, we need hard X-ray observations.

In order to estimate the height difference between white-light and hard X-ray emission, and the effect of the accelerated electrons on the white-light emission in detail, we need to compare hard X-ray images with \textit{Hinode}/SOT white-light images.
As mentioned earlier, \textit{RHESSI} hard X-ray data are not available for this flare.
If the hard X-ray source was located at a height of $600\,{\rm km}$ from the photosphere, as in \citet{Kontar2008}'s event, the height difference between the white-light and hard X-ray emission would be approximately a few hundred km, considering that the continuum emission we observe is distributed over $\sim 400\,{\rm km}$.
If the observed continuum emission is located a few hundred km higher than the $\tau = 1$ photosphere; i.e. around the location of the blue limb, then the emission height of the continuum emission would be very close to the height of hard X-ray emission.
This is almost the same result as \citet{Oliveros2012}.
We can expect that in the near future, hard X-ray images will be observed concurrent with white-light flares, and we can then perform detailed analysis of the height differences between white-light and hard X-ray emissions during flares.

\acknowledgments
\textit{Hinode} is a Japanese mission developed and launched by ISAS/JAXA, with NAOJ as domestic partner and NASA and STFC (UK) as international partners. It is operated by these agencies in co-operation with ESA and NSC (Norway).
A part of this work was carried out with the support of a joint research program grant of the Solar-Terrestrial Environment Laboratory, Nagoya University.
K.~Watanabe's work was supported by the Grant-in-Aid program of the Japan Society for the Promotion of Science Fellows.
We thank E.~Kontar, J.~Oliveros and H.~Hudson for helpful discussions of white-light flare observations.
We also thank A.~Sterling and D.~Brooks for critically reading the paper.

\begin{table}
\begin{center}
\caption{Averaged over the area of $\ge 3\sigma$ intensity of white-light emission of each color region. \label{tbl1}}
\begin{tabular}{crrr}
\tableline\tableline
Observed time & Red region [K] & Green region [K] & Blue region [K]\\
\tableline
18:21:46-18:21:52\,UT & 4822 & 4472 & 4793\\
18:22:05-18:22:11\,UT & 4952 & 4997 & 4976\\
18:22:24-18:22:30\,UT & 4965 & 5020 & 5033\\
18:22:43-18:22:50\,UT & 4990 & 5036 & 5058\\
\tableline
\end{tabular}
\end{center}
\end{table}

\end{document}